\RequirePackage{fix-cm}
\documentclass[smallcondensed]{svjour3}     % onecolumn (ditto)
\smartqed  % flush right qed marks, e.g. at end of proof
\usepackage{graphicx,color}
\usepackage[caption=false]{subfig}
\usepackage[normalem]{ulem}
\usepackage{amsmath}
\usepackage{amssymb}
\usepackage[dvipsnames]{xcolor}
\usepackage[autostyle]{csquotes}
\usepackage{lineno}
\usepackage{hyperref}
\definecolor{Orange}{rgb}{1.00,0.55,0.00}
\usepackage{xcolor}
 \journalname{Frontiers in Mathematics}
\begin{document}

\title{Transmissibility in Interactive Nanocomposite Diffusion: The Nonlinear Double-Diffusion Model
}
\subtitle{Transmissibility in Interactive Nanocomposite Diffusion}


\author{${}^{1}$Amit K Chattopadhyay \and
        ${}^{2}$Bidisha Kundu \and
        ${}^{3}$Sujit Kumar Nath\and
        ${}^{4}$Elias C Aifantis
}


\institute{${}^{1}$A. K. Chattopadhyay \at
Department of Mathematics, Nonlinearity and Complexity Research Group, Aston University \\
Birmingham B4 7ET, UK\\
\email{a.k.chattopadhyay@aston.ac.uk}
\and
${}^{2}$B. Kundu 
\at School of Life Sciences, University of Lincoln \\ Lincoln LN6 7TS, UK \\
\email{bidishakundu.iisc@gmail.com}
\and
${}^{3}$S. K. Nath\at
School of Computing, University of Leeds, Leeds LS2 9JT, UK\\
Faculty of Biological Sciences, University of Leeds, Leeds LS2 9JT, UK\\
\email{s.k.nath@leeds.ac.uk} 
\and
${}^{4}$E. C. Aifantis\at
Laboratory of Mechanics and Materials, Aristotle University of Thessaloniki \\GR-54124 Thessaloniki, Greece\\
\email{mom@mom.gen.auth.gr}
}

\date{Received: 31 December 2021 / Accepted: date}

\maketitle

%\linenumbers

\begin{abstract}

Model analogies and exchange of ideas between physics or chemistry with biology or epidemiology have often involved inter-sectoral mapping of techniques. Material mechanics has benefitted hugely from such interpolations from mathematical physics where dislocation patterning of platstically deformed metals \cite{wa1,wa2,wa3} and mass transport in nanocomposite materials with high diffusivity paths such as dislocation and grain boundaries, have been traditionally analyzed using the paradigmatic Walgraef-Aifantis (W-A) double-diffusivity (D-D) model \cite{eca1,eca2,eca3,akc-eca1,akc-eca2,akc-eca3}. A long standing challenge in these studies has been the inherent nonlinear correlation between the diffusivity paths, making it extremely difficult to analyze their interdependence. Here, we present a novel method of approximating a closed form solution of the ensemble averaged density profiles and correlation statistics of coupled dynamical systems, drawing from a technique used in mathematical biology to calculate a quantity called the {\it basic reproduction number} $R_0$, which is the average number of secondary infections generated from every infected. We show that the $R_0$ formulation can be used to calculate the correlation between diffusivity paths, agreeing closely with the exact numerical solution of the D-D model. The method can be generically implemented to analyze other reaction-diffusion models.


\end{abstract}

\section{Introduction}
\label{intro}
Transport of mass, heat or electricity in inhomogeneous media has been modeled \cite{wa1,wa2,wa3} involving distinct conducting paths such as diffusion in metals containing a large number of dislocations and/or grain boundaries \cite{eca1}, fluid flows in fissured rocks and media with double porosity \cite{eca2,eca3}, heat or electricity conduction in fiber reinforced composites \cite{eca4} have been addressed by Aifantis through continuous models, typically based on coupled sets of linear partial differential equations (the double diffusivity or D-D model) involving 4 phenomenological constants: 2 diffusion coefficients for each one of the two paths and two mass exchange constants between the paths. The above two-state idea was also utilized later in developing the first dynamical model of dislocation patterning, commonly known as the Walgraef-Aifantis (W-A) model \cite{wa1,wa2} that could distinguish between two dislocation populations: slow or “immobile” dislocation and fast or “mobile” ones that brings plastic deformation about. It turns out that the linearized version of the W-A model is identical in form to the D-D model variant of the two-state reaction-diffusion (R-D) formulation used to describe transport of multiple families of species such as vacancies and interstitials in crystalline lattices, impurity, segregation in dislocation and grain boundaries or trapping and precipitation process in alloys.

Over the last two decades, the D-D and W-A models have become quite popular in both the applied mathematics \cite{wa4} and the material science \cite{wa5} communities due to the interesting mathematical properties of the former and robust interpretation of experimental observations of the latter groups of models. This includes implementation of D-D type models in interpreting molecular and mesoscopic transport in condensed matter and cosmological systems \cite{wa6}, {\it i.e.} Most of such models though overlooked the contribution of stochastic forcing or spatial randomness, e.g. surface impurities in materials, restricting the implementation of such models in explaining experimental observations. A recent series of stochastically driven D-D models \cite{akc-eca1,akc-eca2,akc-eca3} have not only addressed this issue of steering qualitative phenomenological models closer to experimental descriptions, technically these have opened up further possibilities of cross-disciplinary implementation of these models from material science to other fields and vice versa. 

{\color{black}{Anomalous diffusion involving multiple species and media has for long remained an interesting field of research. The diffusive behaviour changes their characteristics depending on the choice of medium. There are many plausible reasons for this. One of them is that there exists some void grain boundary in the medium which represents an especially high diffusivity path inside the medium. Also, there are relatively narrow domains in the medium where the diffusion rate is slower. Simultaneous diffusion in multiple media has been traditionally analyzed using double-diffusion models \cite{eca1,eca2,eca3,eca4,eca5} that is a coupled system of partial differential equations involving interacting variables. These D-D models considered two species of diffusive elements, one that follows the regular path, another following a high-diffusive path, with eventual dynamics determined by a dynamical equilibrium of these competing paths.}}

\subsection{State of the Art} 
Elias C. Aifantis developed and introduced the concept of double diffusion step by step in \cite{aifantis1979continuum,aifantisprb1979}-\cite{eca2,eca3}. A continuum basis for diffusion in regions with multiple diffusivity was introduced in \cite{aifantis1979continuum}. Simultaneously, in \cite{aifantis1979new}, the diffusion in media with a continuous distribution of high-diffusivity paths was modelled. Finally, Aifantis provided a formulation generalising this idea of the diffusion in solid media for wide range of applicability in different physical process, in double porosity \cite{wilson1982}, from metallurgy to soil science \cite{eca5} polymer physics and geophysics, in \cite{eca2,eca3}. Another explanation of this double diffusion model was provided in \cite{hill1980discrete} using the concept of discrete random walk model. In \cite{eca2,eca3}-\cite{hill1980theory}, Aifantis and Hill studied the basic mathematical questions of the model. Mainly they studied uniqueness, maximum
principles and basic source solutions in  \cite{aifantis1980problem,hill1980theory} and boundary value problems in \cite{hill1980theory}. Kuttler and Aifantis studied the existence and uniqueness of the weak form of the nonclassical diffusion equation in \cite{kuttler1987existence}.

The diffusion process in a media is not deterministic. Indeed there are stochastic effects initiated and controlled by several factors. Randomness can be related to thermal fluctuations, grain size changes, impurity effects, etc. {\color{black}{Recent studies of a type of non-equilibrium system involving multiple states of diffusion of a diffusing species, called stochastic resetting, is governed by a dynamics similar to a double-diffusion system \cite{sujitBrMo}.}} These types of interactive features play an important role in the process and it became necessary to take into account of the stochastic agents. In nanoscales or nanopolycrystals, the diffusion near the grain boundary following two paths, regular and high diffusive, can be affected by stochastic fluctuations \cite{akc-eca1}. Deterministic internal length gradient method can not completely explain relaxation time for diffusion in nanopolycrystals. Considering boundary layer fluctuations, stochasticity was added in the modeling and first stochastic gradient nanomechanics (SGNM) model was proposed in \cite{akc-eca2}. Using SGNM model, relaxation time is discussed thoroughly for a specific superconductors \cite{konstantinidis2001application} in \cite{akc-eca3}. Also, linear stochastic resonance has been predicted and how stochastic effects start affecting the system is explained in \cite{akc-eca3}.

\subsection{Open Questions}
		
The present article is the first in line to provide a closed form approximate solution of the nonlinear model resulting from a combination of the D-D and W-A models. The D-D:W-A composite model leads to a coupled system of reaction-diffusion (R-D) equations where the diffusion terms are identical to those contained in both models, the linear terms as in the D-D model \cite{akc-eca1,akc-eca2} while the nonlinear terms resemble the W-A model (3rd order), but contain, in addition, cross-coupled terms (2nd order). The underlying physical picture represents a system of multiple diffusive relaxation, including boundary layer shear (nonlinear terms), and driven by stochastic forcing as in the two models. We consider simultaneous diffusion in the lattice or grain interior along the grain boundaries but allow now for trapping effects. In other words, diffusing species can be trapped in both grain interiors along dislocation cores and dislocation dipoles as well as in the counterparts of the defects within the grain boundary space. 
	
\par	
While diffusion mediated interaction between multiple species is intrinsically nonlinear, accounting for sample impurity, reaction-diffusion terms and the like, traditional analyzes have relied on exactly solvable linear models, both in deterministic and stochastic regimes. While there is no paucity of numerical estimation of nonlinear models, including those for double diffusion (\cite{akc-eca2} and references therein), from the perspective of theoretical modeling, analytical clarity had to give way to quantitative precision. More importantly, a closed form solution offers a direct methodological link between the process and its parameters that is not available from brute numerical evaluation in a multi-parameter space. The present study is thus a major breakaway from the linearity assumption, retaining closer proximity to experiments while also comparing favorably with numerical solutions, as we will later show. 

\par
Nonlinear diffusion equations, a classical example of parabolic type equations, play an important role in the modeling of different phenomena. Broadly speaking, one may refer to, two classes of diffusion equations with nonlinearity \cite{vazquez2017mathematical}. One is for free boundary problems such as the distribution of temperature in a homogeneous material during phase-transition \cite{rubinvsteuin2000stefan}, i.e., the time evolution of the phase boundaries, the so called Stefan problem. 
 
 \par
 Another is for reaction-diffusion problems, such as the Fisher-Kolmogorov-Petrovsky-Piskunov (Fisher-KPP) model for propagation of an advantageous gene in a population \cite{el2019revisiting}, \cite{murraybook}, Gray Scott Model  for diffusion of chemical species \cite{pearson1993complex}. Similar style of modeling can be found for dislocation profiles in a material.  Walgraef and Aifantis (WA model) proposed a model of a system of Reaction-Diffusion equations considering two profiles of dislocation flows, immobile dislocation for slow moving and mobile for relatively speedy moving \cite{wa1}. The W-A model has been studied extensively numerically, towards bifurcation analysis and pattern formation in \cite{wa4}, \cite{wa5}. None of these were targeting a closed form analysis, as is the target in this study.

\par
Following a general introduction and pointers to open questions in Section 1, Section 2 summarises the model equation and provides a physical explanation of the mechanisms involved. Section 3 represents the nondimensionalized representation of the model in Section 2, the relevant non-dimensional governing equations and their linear stability analysis. Section 4 first discusses a popular method used in mathematical epidemiology in the calculation of the time varying reproduction number $R_0$, then identifies the phenomenology as one of analysing the covariance of two (or more) coupled variables in a dynamical system, and then uses this hypothesis to connect with the cross-coupling between D-D variables. Section 5 provides the anatomy of the time ($t$) evolution of the reproduction number $R_0(t)$ in the equivalent D-D model as one measuring the strength of cross-correlation between the different diffusing species. Finally, Section 6 summarizes the outcomes of this continuum (approximate) mapping that then is compared against direct numerical evaluation of this model.		

\section{The Model}
In this work, we focus on a closed-form, albeit approximate, solution of the D-D model for nano polycrystal diffusion by considering nonlinear source terms in the original W-A model, representing additional exchange of diffusion species between the two paths. These new non-linear exchange terms represent the transfer of diffusion species through dislocation atmosphere i.e., diffusing species segregated in dislocation cores and dislocation dipoles. We study in particular how the "transmissibility" of the species affect their diffusion and corresponding trapping processes.
We observe how their internal interactions can affect their behaviour. We study how the transmissibility of the species affects their diffusion. 

Considering $\tilde{\rho_{1}}$ and $\tilde{\rho_{2}}$ as the concentrations/densities for the two distinct D-D species along two different paths, the governing equations of diffusion are given by

\begin{subequations}
\begin{eqnarray}
 \frac{\partial \tilde{\rho_{1}}}{\partial \tilde{t}}=D_{1} \frac{\partial^{2} \tilde{\rho_{1}}}{\partial \tilde{x^{2}}}-k_{1} \tilde{\rho_{1}}+k_{2} \tilde{\rho_{2}}+ 	\lambda_{1} \tilde{\rho_{1}}\tilde{\rho_{2}}+\sigma_{1} \tilde{\rho_{1}}^{2} \tilde{\rho_{2}}
 \label{modelequation1}
  \\
  \frac{\partial \tilde{\rho_{2}}}{\partial\tilde{ t}}=D_{2} \frac{\partial^{2} \tilde{\rho_{2}}}{\partial\tilde{ x^{2}}}+k_{1} \tilde{\rho_{1}}-k_{2} \tilde{\rho_{2}}+\lambda_{2} \tilde{\rho_{1}}\tilde{\rho_{2}}+\sigma_{2} \tilde{\rho_{1}} \tilde{\rho_{2}}^{2}
  \label{modelequation2}
\end{eqnarray}
\end{subequations}

where $D_{1}, D_{2}$ are diffusion coefficients, $k_{1}, k_{2}$ are the rate mass exchange between different paths. The nonlinear terms represent the 
interactions between different species (or dislocation paths, in case of two diffusive paths in the material body) when the density 
of one species influences the creation or annihilation of the 
other. 

\section{Non-Dimensionalization of the Double Diffusing Walgraef-Aifantis model}

Let $\tilde{x}=ax$, $\tilde{t}=b t,  \tilde{\rho_{1}}=c_{1} \rho_{1}, \tilde{\rho_{2}}=c_{2} \rho_{2}$. Substituting in Eqs.  (\ref{modelequation1}-\ref{modelequation2}), then assuming that the diffusion coefficients remain unchanged after scaling, and choosing coefficients of nonlinear product terms as unity after scaling, for $\sigma_1=\sigma_2=\sigma$, we have 
\begin{subequations}
\begin{eqnarray}
 \frac{\partial \rho_{1}}{\partial t}=D_{1} \frac{\partial^{2} \rho_{1}}{\partial x^{2}}-\left(\frac{k_{1}\sigma} {\lambda_{1} \lambda_{2}}\right)\rho_{1}+\left(\frac{k_{2} \sigma}{\lambda_{1}^{2}}\right) \rho_{2}+ \rho_{1}\rho_{2}+ \rho_{1}^{2} \rho_{2} 
  \label{Scalemodelequation1}
  \\
  \frac{\partial \rho_{2}}{\partial t}=D_{2} \frac{\partial^{2} \rho_{2}}{\partial x^{2}}+
\left(\frac{k_{1}\sigma} {\lambda_{2}^{2}}\right)\rho_{1}-\left(\frac{k_{2} \sigma}{\lambda_{1} \lambda_{2}}\right) \rho_{2}+\rho_{1}\rho_{2}+\rho_{1} \rho_{2}^{2}.
 \label{Scalemodelequation2}
\end{eqnarray}
\end{subequations}
{\color{black}{Note, the variables $\rho_1$ and $\rho_2$, representing Eqns. (\ref{Scalemodelequation1},\ref{Scalemodelequation2}), are non-dimensional. The numerical model uses this system of a non-dimensional dynamical system.}} 

\subsection{Linear Stability Analysis}
Equations (\ref{Scalemodelequation1},\ref{Scalemodelequation2}) can be represented as the following coupled reaction-diffusion model
\begin{subequations}
\begin{eqnarray}
 \frac{\partial \rho_{1}}{\partial t}=D_{1} \frac{\partial^{2} \rho_{1}}{\partial x^{2}}+F_{1}( \rho_{1},  \rho_{2})
 \label{newmodel1}
   \\
  \frac{\partial \rho_{2}}{\partial t}=D_{2} \frac{\partial^{2} \rho_{2}}{\partial x^{2}}+F_{2}( \rho_{1},  \rho_{2}),
  \label{newmodel2}
\end{eqnarray}
\end{subequations}
where
$
F_{1}( \rho_{1},  \rho_{2})=-\left(\frac{k_{1}\sigma} {\lambda_{1} \lambda_{2}}\right)\rho_{1}+\left(\frac{k_{2} \sigma}{\lambda_{1}^{2}}\right) \rho_{2}+ \rho_{1}\rho_{2}+ \rho_{1}^{2} \rho_{2}$, and
$
F_{2}( \rho_{1},  \rho_{2})=\left(\frac{k_{1}\sigma} {\lambda_{2}^{2}}\right)\rho_{1}-\left(\frac{k_{2} \sigma}{\lambda_{1} \lambda_{2}}\right) \rho_{2}+\rho_{1}\rho_{2}+\rho_{1} \rho_{2}^{2}$. We analyze the system stability near the {\it Homogeneous Equilibrium} (HE) state  or at the uniform steady state $(\rho^{*}_{1}, \rho^{*}_{2})$ \cite{wa1}, \cite{wa5}, i.e., where 
\begin{subequations}
\begin{eqnarray}
F_{1}(\rho^{*}_{1}, \rho^{*}_{2})&=&0,
\\
F_{2}(\rho^{*}_{1}, \rho^{*}_{2})&=&0
\end{eqnarray}
\end{subequations}
Solving these equations we get the HE state, $(\rho_1^*,\rho_2^*) = (0,0)$. Perturbing around this equilibrium state, perturbations defined as $(\hat{\rho_1},\hat{\rho_2})$, we get $\rho_{1}=\rho^{*}_{1}+\hat{\rho_1}, \:\rho_{2}=\rho^{*}_{2}+\hat{\rho_2}$.

Now near the HE states, the linearised version of the Eqs (\ref{newmodel1}-\ref{newmodel2}) are 

\begin{eqnarray}
 \frac{\partial \overline{\rho}}{\partial t}=\begin{pmatrix}
D_1& 0\\
0 & D_2
\end{pmatrix}\frac{\partial^{2}\overline{\rho}}{\partial x^{2}}+ J_{F}\overline{\rho}
 \label{linearmodel}
 \end{eqnarray}
where $\overline{\rho}=\begin{pmatrix}
  \hat{\rho_1}\\ 
  \hat{\rho_2}
\end{pmatrix}$, and
$J_{F}$ is the Jacobian of $\begin{pmatrix}
  F_{1}( \rho_{1},  \rho_{2})\\ 
F_{2}( \rho_{1},  \rho_{2})
\end{pmatrix}$
at the equilibrium states $(\rho^{*}_{1}, \rho^{*}_{2})$. We consider $\overline{\rho}= \begin{pmatrix}
\phi_1(t)\\ 
  \phi_2(t)
\end{pmatrix} e^{i\omega x}$ for real $\omega$ and get 

\begin{eqnarray}
\begin{pmatrix}
\frac{d\phi_1}{dt}\\ 

 \frac{d\phi_2}{dt}
\end{pmatrix}=\left(J_{F}-\omega^{2}D\right)\begin{pmatrix}
\phi_1(t)\\ 
  \phi_2(t)
\end{pmatrix},
  \end{eqnarray}
where $D=\begin{pmatrix}
D_1& 0\\
0 & D_2
\end{pmatrix}$.
As this is a system of two variables, the signatures of trace and determinant of the matrix $\left(J_{F}-\omega^{2}D\right)$ defines the stability of the system.  The determinant should be always positive and trace should be always negative for all real values of $\omega$. We test these conditions for the HE state at $(0,0)$ and arrive at the following closed form expressions for the Trace (Tr) and Determinant (Det) of the model:

\begin{subequations}
\begin{eqnarray}
 \text{Tr}\left(J_{F}-\omega^{2}D\right)=-D_{1} \omega ^2-D_{2} \omega ^2-\frac{k_{1} \sigma
   }{\lambda_{1} \lambda_{2}}-\frac{k_{1} \sigma
   }{\lambda_{1} \lambda_{2}}
   \\
\text{Det} \left(J_{F}-\omega^{2}D\right)=D_{1} D_{2} \omega ^4+\frac{D_{1} k_{2} \sigma 
   \omega ^2}{\lambda_{1} \lambda_{2}}+\frac{D_{2}
   k_{1} \sigma  \omega ^2}{\lambda_{1} \lambda_{2}}
\end{eqnarray}
\end{subequations}
 For $D_{1} >0, D_{2} >0, k_{1}>0, k_{2}>0,\sigma >0, \lambda_{1} >0 \lambda_{2} >0,\omega^{2} >0 $, 
 Trace is always negative and determinant is always positive. Hence, the HE state at $(0,0)$ is a stable state.

\section{Biology to Materials' Modeling}
It is usual practice in infectious disease epidemiology and modeling to measure the \enquote{speed} of the propagation of the infection. This measurement is generally called the {\it basic reproduction number} $R_{0}$ that effectively equates to the number of secondary infections generated from each infected member of the population. $R_{0}$ depends on the numbers of currently infected, susceptible and the rate of infection in the population. This $R_0$ is the threshold parameter for an infectious disease, determining whether it becomes an epidemic, pandemic, or extinct in a community. The epidemiologists follow several methods to calculate $R_{0}$. Two of these, referred to as the {\it next generation method} and the {\it age-structured method} \cite{R0-Nishiura,R0-Ferguson} are widely used. Both are effective and popular in infection modeling studies. 

In this work, we show how the concept of $R_{0}$ can be made an auxiliary method in studying  the diffusion process in a medium. We show how the profile of time evolution of $R_{0}$ can help us to understand the diffusion-dynamics of two species, and can be used as a substitute to the enumeration of correlation functions. 

{\color{black}{Our starting point in this \enquote{reverse mapping} scheme from mathematical biology to material science relates to the origin of the concept of {\it basic reproduction number} $R_0$. Let $I(t)$ be a time-evolving quantity whose value at time $t$ is dependent on its values at
previous time points. This essential non-Markovian distribution ensures non-trivial values for all $I(t-\tau)$, where $0\leq\tau\leq t$, as long as $I(t)$ is defined. Representing $I(t)$ as the number of infected individuals
at time $t$ in a population, non-Markovian kinetics ensures that $I(t)$ should
depend on the number of infected, present in the population, at
time $t-\tau$, since the new infections can only be generated by
the previous infections. The time required for an infected individual
to generate a new infection, from the onset of its infection,
is called the generation time $\tau$. Clearly, $\tau$ is a non-negative
continuous random variable which has a probability density function,
say $g(\tau)$. In the case of infectious diseases, $g(\tau)$ is generally taken
as Gamma or a lognormal distribution.}}

\section{\enquote{Generation time} in Double Diffusion: Comparison with random walk model}
We can think of double diffusion as the continuum limit 
of a random walk model where the random walker diffuses 
along two different diffusive paths, occasionally jumping 
between them \cite{hill1980discrete}. Here, by different 
diffusivity of paths we mean the probability of left jump ($p_i$), 
right jump ($q_i$), staying at same position without making 
any jump ($r_i$) are different for the two paths, where $i=1, 2$ 
for path-1 and path-2. Let us introduce a random time interval 
$\tau$, having a probability density function $g(\tau)$, during 
which the walker diffuses along the same path before making any 
jump to the other path. The time $\tau$ can be thought of as 
the generation time for the double diffusion model, in parallel 
with the well defined generation time for an infectious disease. 
For this random walk model of double diffusion, let the probability 
of jumping from path-1 to path-2 be $\mathbb{P}(1\to 2)=s_1$ and 
same for path-2 to path-1 be $\mathbb{P}(2\to 1)=s_2$. Therefore, 
the generation time in case of our double diffusion model 
can be compared to the survival time of the random walker on a 
single path before making a jump to the other. Now, for the random 
walk model of double diffusion we must have $p_i+q_i+r_i+s_i=1$ 
for $i=1, 2$. Therefore, the probability that the random walker 
continues on path $i$, in two consecutive jumps, is $(1-s_i)$. 
Hence the corresponding survival probability on path $i$ is given 
by a geometric distribution. More explicitly, the probability 
that the walker will stay on path $i$ for $n$ consecutive jumps 
is $(1-s_i)^ns_i$. Motivated by the fact that the geometric 
distribution has memoryless property, and exponential distribution 
is the only continuous distribution having memoryless property, it 
is reasonable to assume that the generation time is exponentially 
distributed in the continuum limit of this random walk model of 
double diffusion.

\subsection{The Reproduction Number $R0$ and its Mapping to the D-D Model}
Following the similar concept in the case of double diffusion, we can think of
the density $\rho(t)$ of certain species is dependent on its density
$\rho(t-\tau)$ at earlier times ($0\leq\tau\leq t$). The time required for generating
new particles of the same species from the old ones is a continuous
random variable $\tau$ with some density function $g(\tau)$. Therefore,
in the same token, as defined in case of infectious diseases, we can
define a quantity $R_0$ for a species in a double-diffusive process as

\begin{eqnarray}
R_0(t) &=& \frac{\rho(t)}{\int_0^{\infty}  \rho(t-\tau) ~ g(\tau) ~ d\tau}.
\end{eqnarray}

where $g(\tau)$ is the generation time. 

 \begin{figure}[ht!]
\captionsetup[subfigure]{position=top}
    \begin{tabular}{ll}
 \subfloat[Autocorrelation and $R_{0}$ of $\rho_{1}$ at $x=0.2$]{\includegraphics[height=5cms,width=5.5cms]{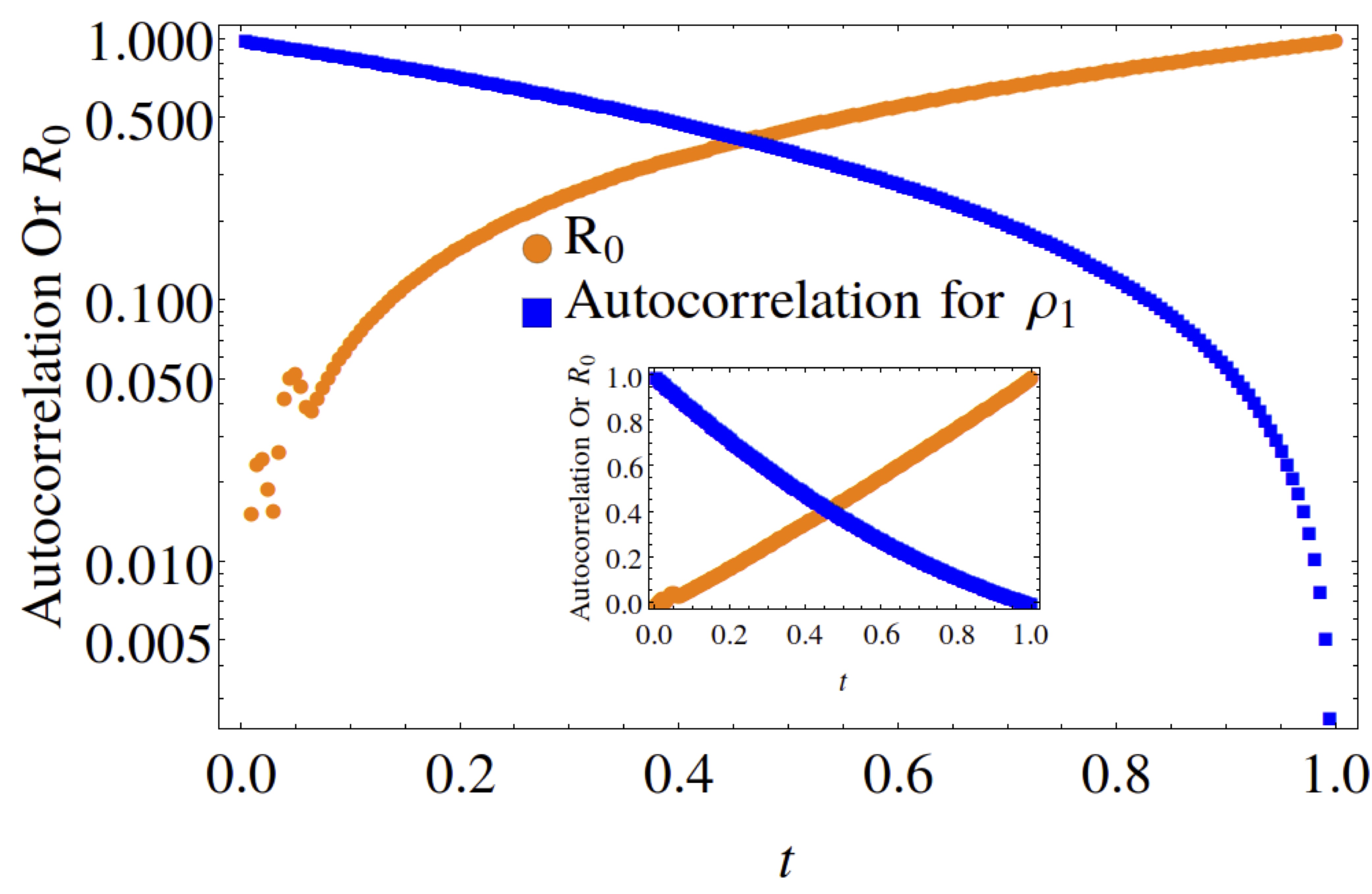}}&
   \subfloat[Autocorrelation and $R_{0}$ of $\rho_{2}$ at $x=0.2$]{\includegraphics[height=5cms,width=5.5cms]{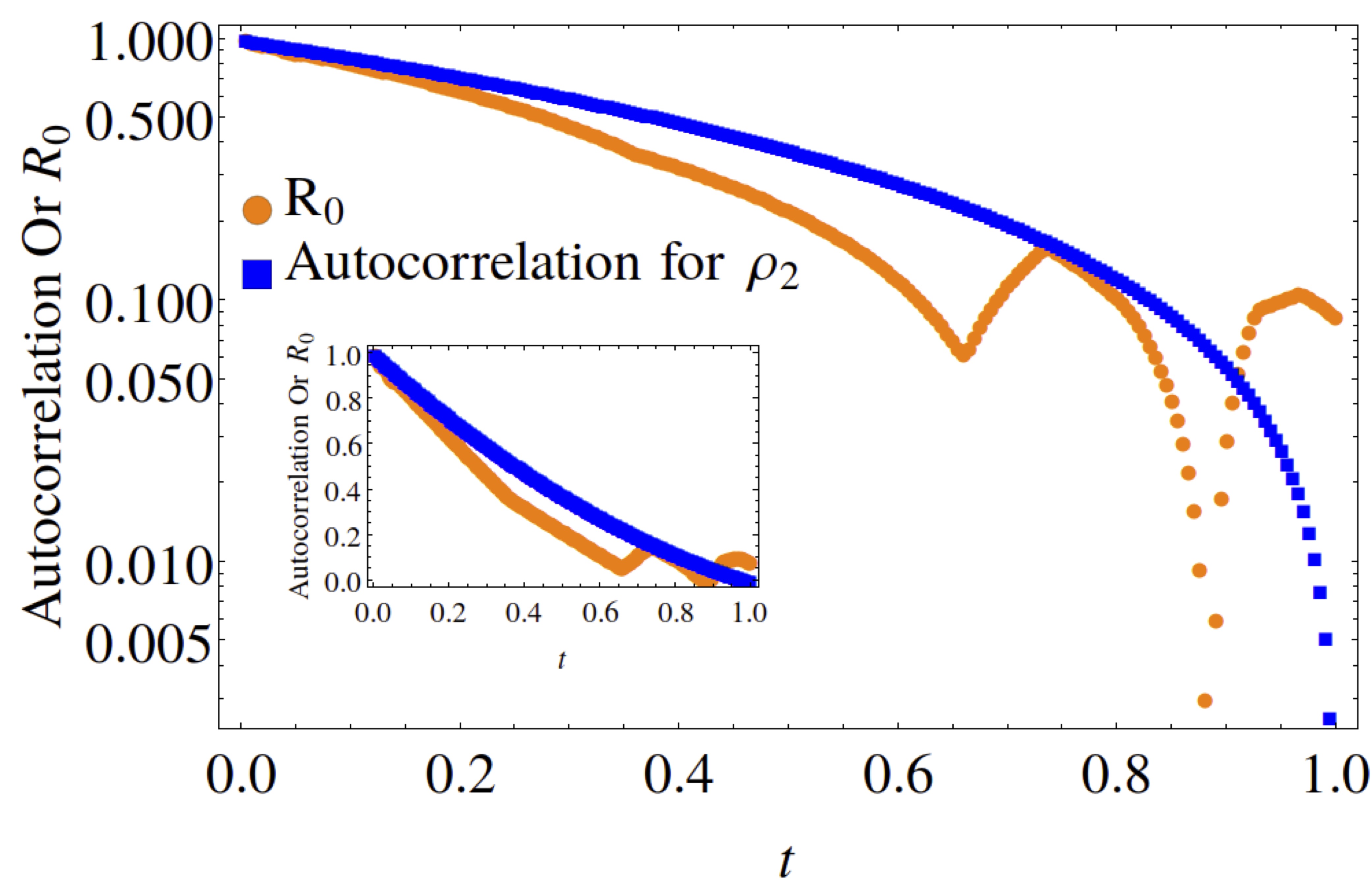}}
      \end{tabular}
        \caption{Comparison of normalized autocorrelation and $R_{0}$ at $x=0.2$. Outset is in log scale and inset is 
        in original scale. Panel (a) shows that the autocorrelation (in blue) is 
        decreasing at later times although the $R_0$ is increasing for the 
        species $\rho_1$ (orange curve). This indicates that at $x=0.2$ the 
        production of $\rho_1$ occurs from its own population at small 
        timescales. On the other hand, in panel (b), the simultaneous decreasing 
        behavior of the autocorrelation and $R_0$ indicate that at $x=0.2$ the 
        production of $\rho_2$ also occurs at smaller timescales. However, 
        in case of $\rho_2$, its production occurs due to the conversion of 
        $\rho_1$ into $\rho_2$, since the measure of self-creation $R_0$ is 
        decreasing in this case.}
        \label{fig0p2}
\end{figure}
%%%%%%%%%%%%%%%%%%%%%%%%%%%%%%%%%
  \begin{figure}[ht!]
\captionsetup[subfigure]{position=top}
    \begin{tabular}{ll}
 \subfloat[Autocorrelation and $R_{0}$ of $\rho_{1}$ at $x=0.3$]{\includegraphics[height=5cms,width=5.5cms]{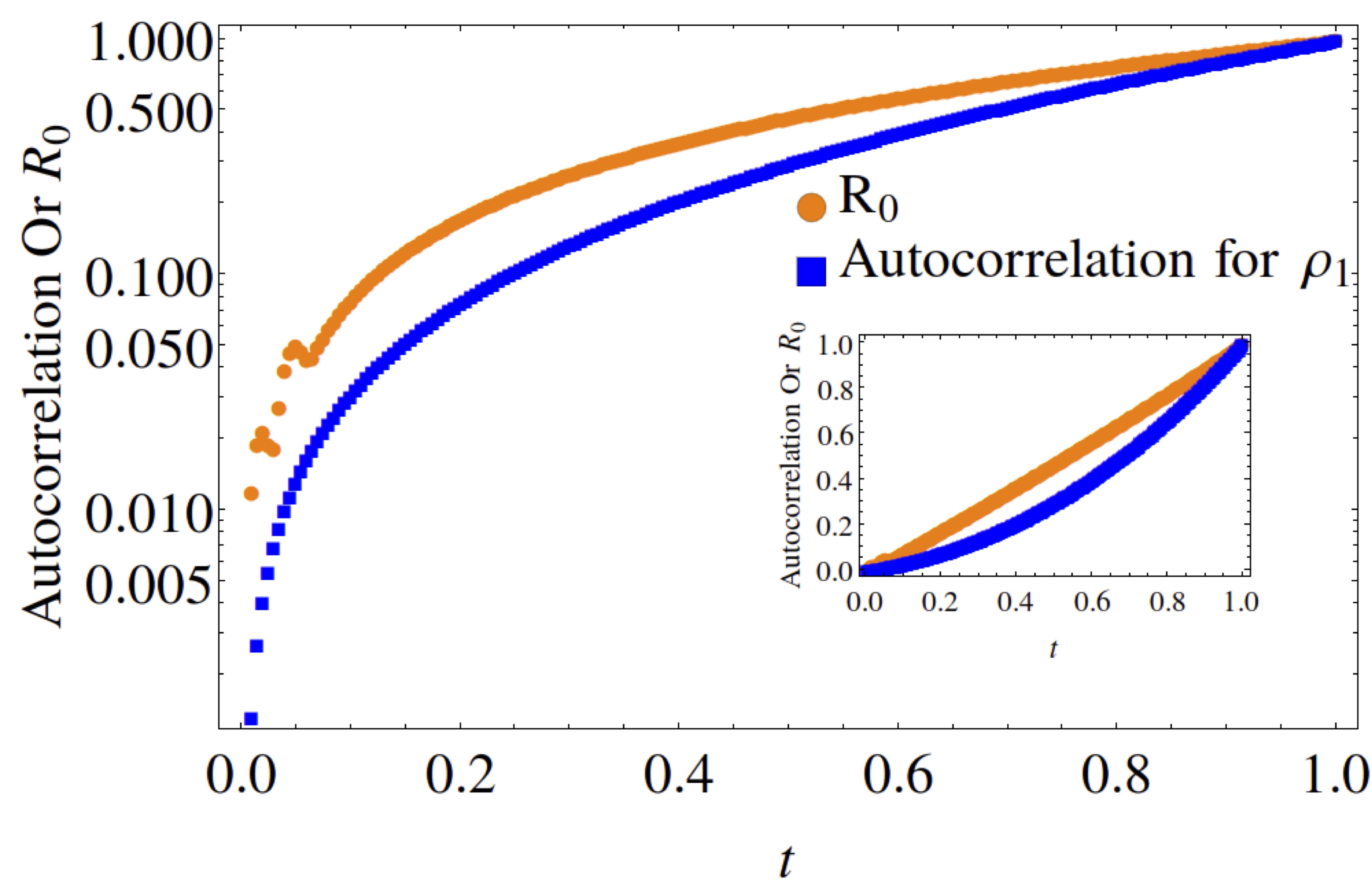}}&
   \subfloat[Autocorrelation and $R_{0}$ of $\rho_{2}$ at $x=0.3$]{\includegraphics[height=5cms,width=5.5cms]{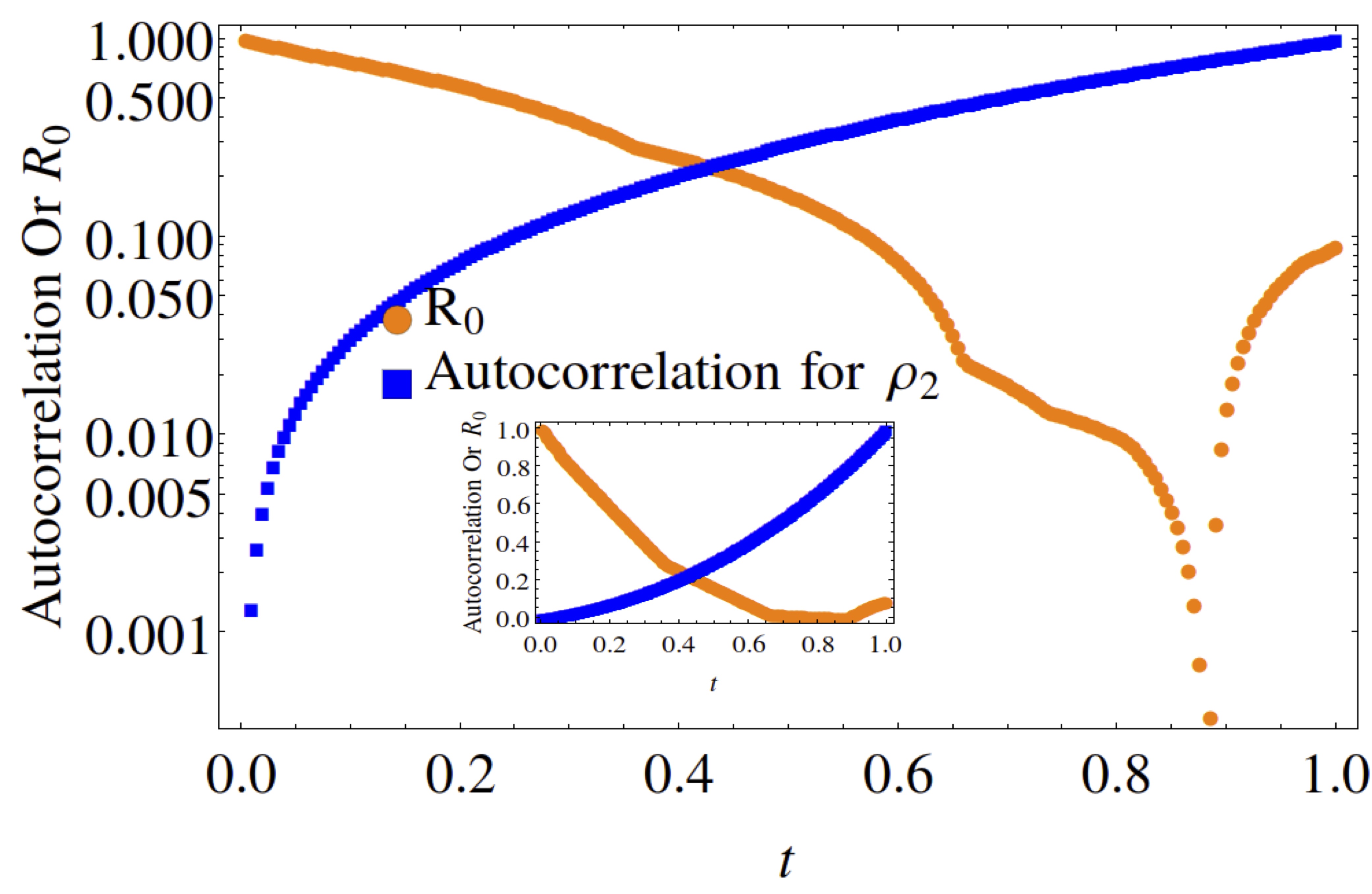}}
      \end{tabular}
        \caption{Comparison of normalized autocorrelation and $R_{0}$ at $x=0.3$. Outset is in log scale and inset is 
        in original scale. Panel (a) shows that the autocorrelation function 
        and $R_0$ for the species $\rho_1$ have the same increasing trend as a 
        function of time. The understanding we get from this trend is that 
        the species $\rho_1$ is produced fom its own species for long time-scales. 
        Panel (b) indicates that the dynamics of the species $\rho_2$ is quite 
        different than $\rho_1$. The species $\rho_2$ is produced from the 
        conversion of $\rho_1$ instead from its own ancestors, as the $R_0$ for 
        $\rho_2$ decreases over time.}
        \label{fig0p3}
\end{figure}

\begin{figure}[ht!]
\captionsetup[subfigure]{position=top}
    \begin{tabular}{ll}
 \subfloat[Autocorrelation and $R_{0}$ of $\rho_{1}$ at $x=0.5$]{\includegraphics[height=5cms,width=5.5cms]{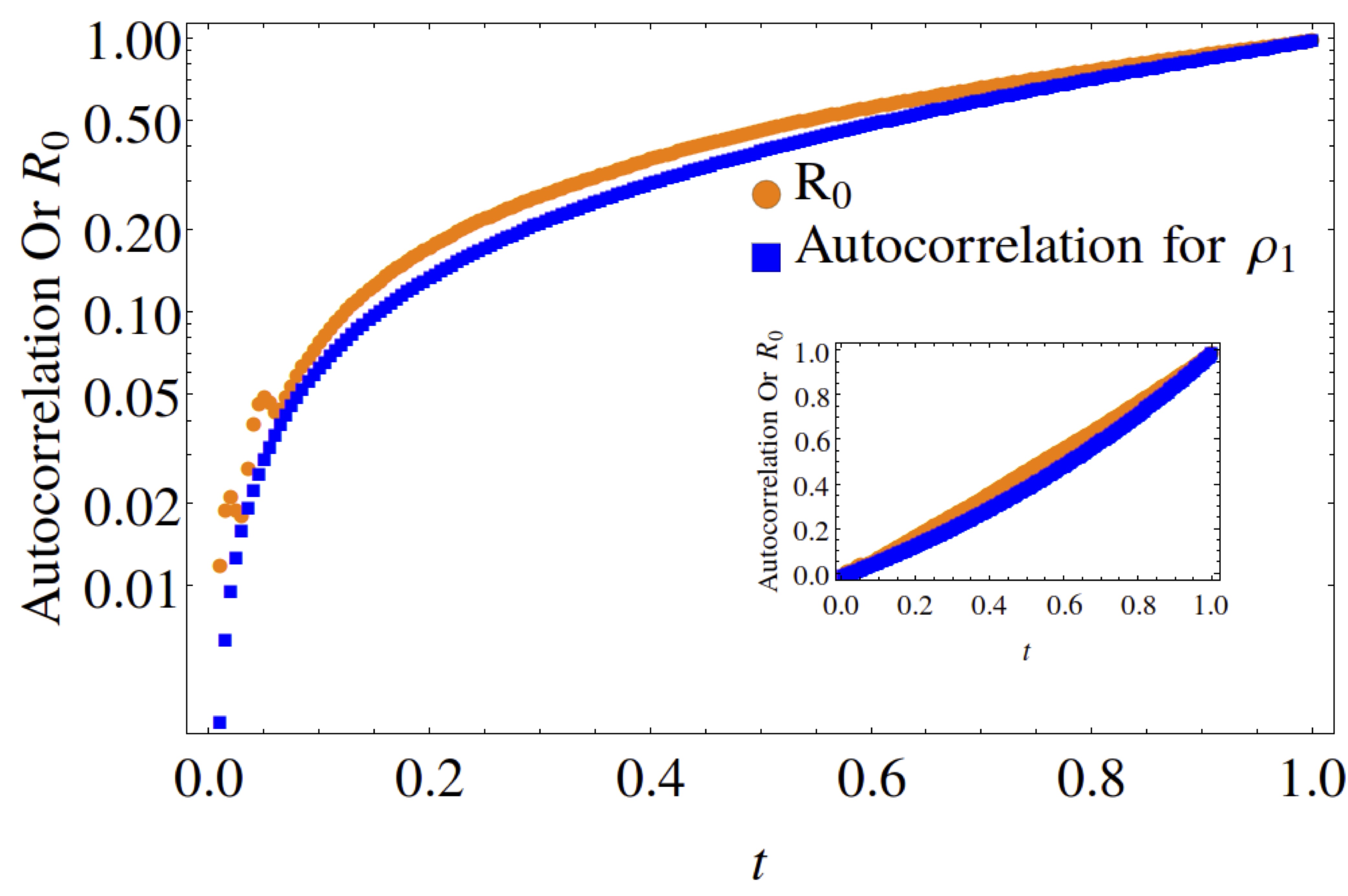}}&
   \subfloat[Autocorrelation and $R_{0}$ of $\rho_{2}$ at $x=0.5$]{\includegraphics[height=5cms,width=5.5cms]{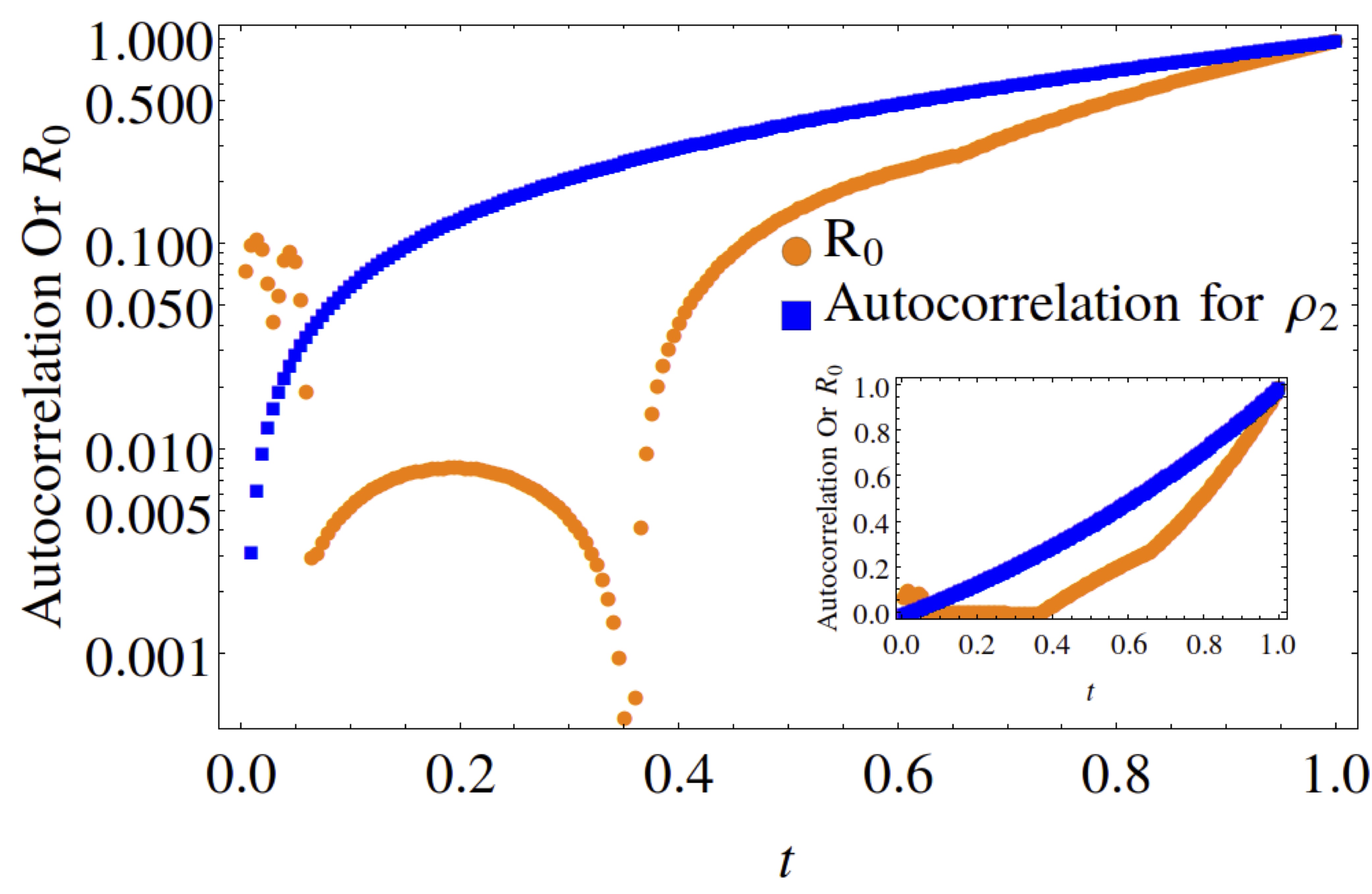}}
      \end{tabular}
        \caption{Comparison of normalized autocorrelation and $R_{0}$ at $x=0.5$. Outset is in log scale and inset is in original scale. Panel (a) shows similar trend for $R_0$ and autocorrelation 
        as in Fig.\ref{fig0p3}(a), which indicates that the production of $\rho_1$ is boosted by 
        the abundance of its ancestor. Whereas, the production of $\rho_2$ occurs both by 
        the conversion of $\rho_1$ for some time and due to its own population at later 
        times.}
        \label{fig0p5}
\end{figure}

\begin{figure}[ht!]
\captionsetup[subfigure]{position=top}
    \begin{tabular}{ll}
 \subfloat[Autocorrelation and $R_{0}$ of $\rho_{1}$ at $x=0.6$]{\includegraphics[height=5cms,width=5.5cms]{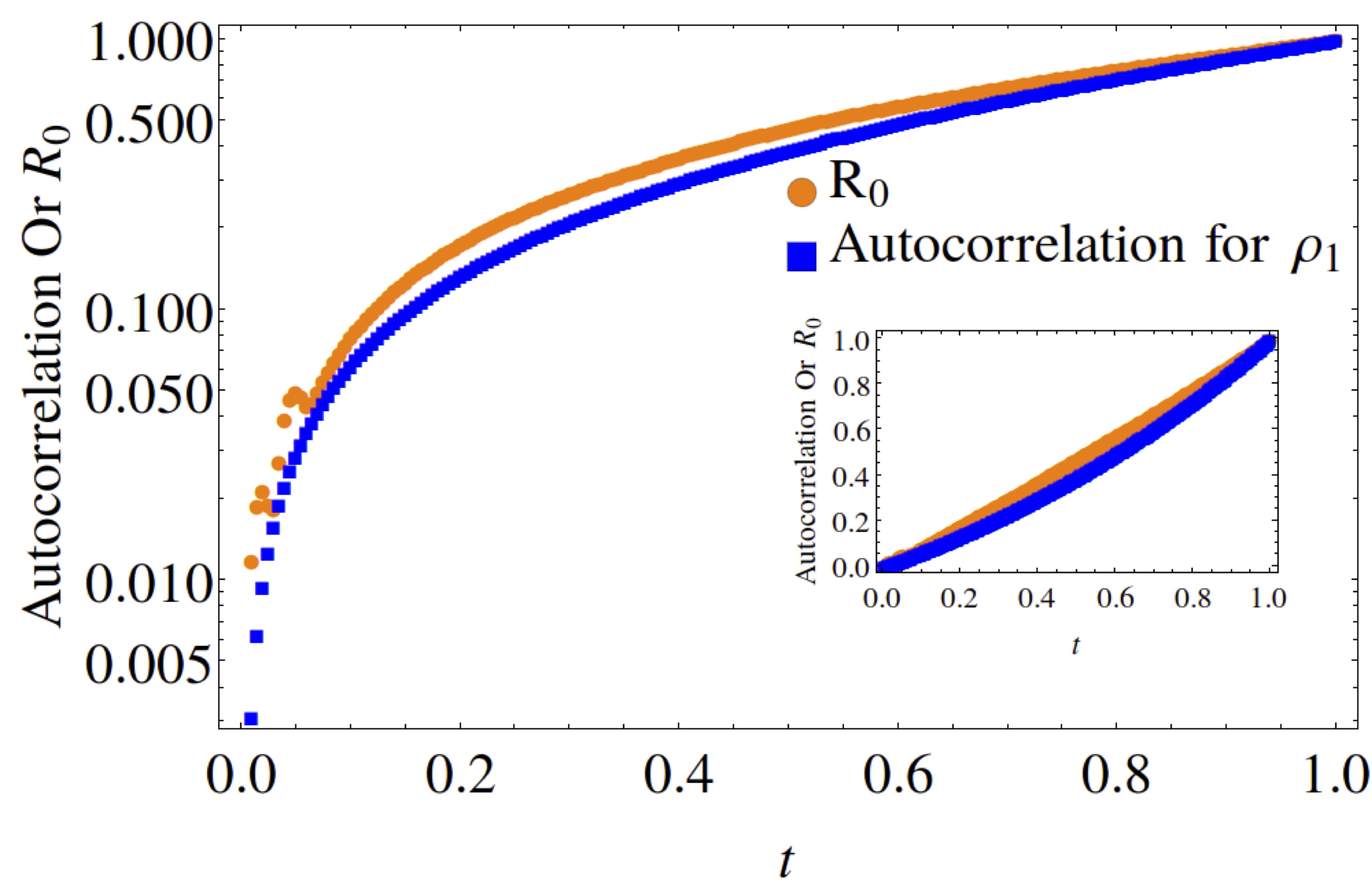}}&
   \subfloat[Autocorrelation and $R_{0}$ of $\rho_{2}$ at $x=0.6$]{\includegraphics[height=5cms,width=5.5cms]{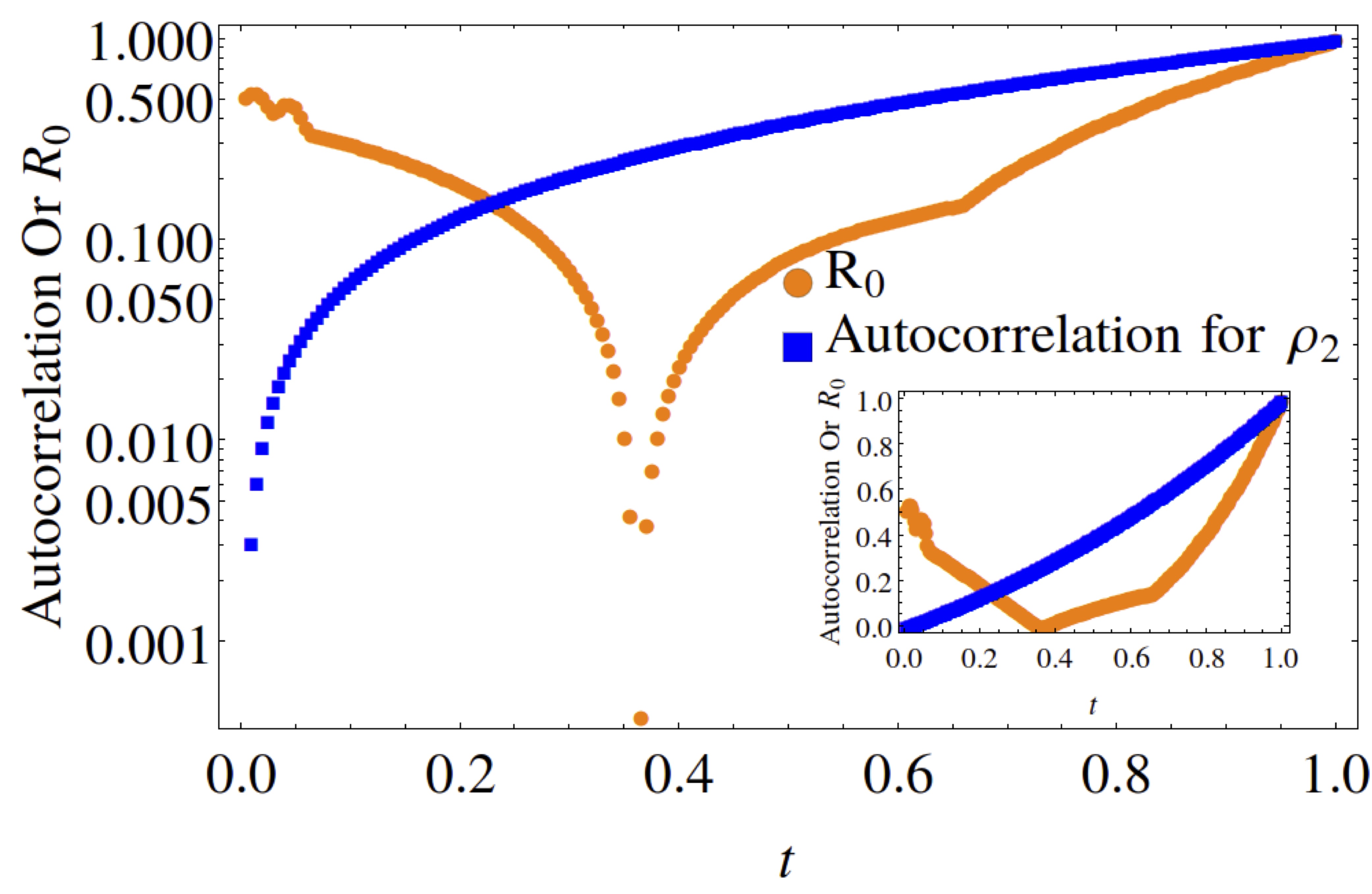}}
      \end{tabular}
        \caption{Comparison of normalized autocorrelation and $R_{0}$ at $x=0.6$. Outset is in log scale and inset is in original scale. 
        The dynamics at $x=0.6$ is almost similar to that of at $x=0.5$, as 
        shown in Fig.\ref{fig0p5}.}
        \label{fig0p6}
\end{figure}

{\color{black}{Figures \ref{fig0p2}, \ref{fig0p3}, \ref{fig0p5}, \ref{fig0p6} compare the $R_0$ equivalent of an equivalent epidemic rate model with 
that of the autocorrelation function (Eqn. (\ref{Scalemodelequation1},\ref{Scalemodelequation2}) at the quantitative level, providing interesting insights into the D-D 
reaction-diffusion model. It can be easily understood from the mathematical 
expression of $R_0(t)$ that it is a measure of the production rate of a 
species from the population of the same species at an earlier epoch, 
rather than due the conversion of other species. On the other hand, autocorrelation 
is a measure of the abundance of a species as a whole, aggregating the production 
of a species from it's own population as well as due to the conversion of 
other species. Therefore, the autocorrelation function together with the 
time-varying $R_0(t)$, gives us interesting spatio-temporal insights about 
the observed abundance}}

{\color{black}{A comparison between Figures \ref{fig0p2} and \ref{fig0p3} with \ref{fig0p5} clearly indicates that while asymmetric cases ($x < 0.5$) ensure only partial convergence between the $R_0$ and autocorrelation profiles, {\it i. e.} only one of the two double diffusing variables match both profiles, at $x=0.5$, the profiles match (approximately) for both variables.}}
%%%%%%%%%%%%%%%%%%%%%%%%%%%%%%%%%
%%%%%%%%%%%%%%%%%%%%%%%%%%%%%%%%%
  \begin{figure}[ht!]
\captionsetup[subfigure]{position=top}
    \begin{tabular}{ll}
 \subfloat[Autocorrelation and $R_{0}$ of $\rho_{1}$ at $x=0.7$]{\includegraphics[height=5cms,width=5.5cms]{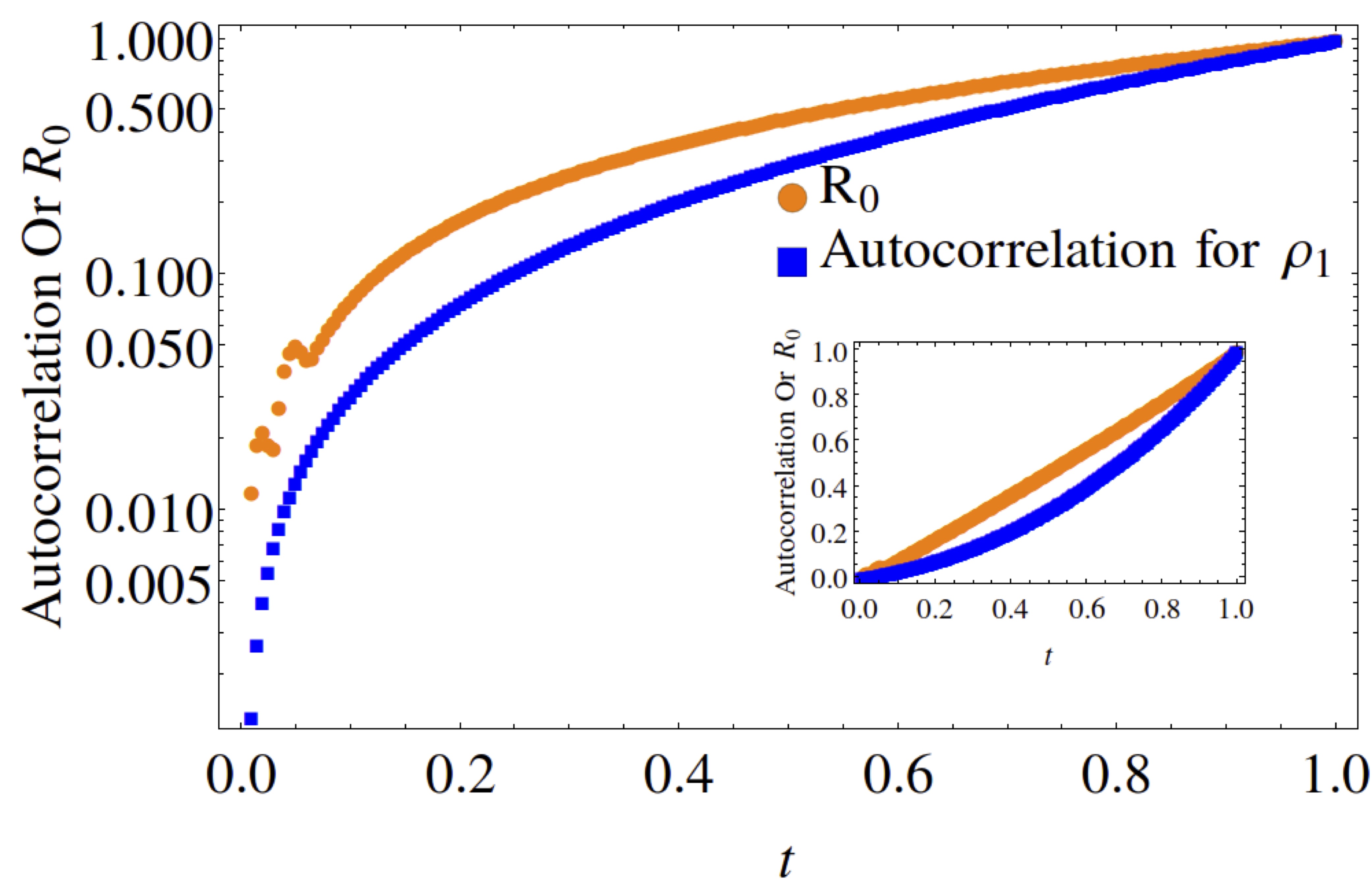}}&
   \subfloat[Autocorrelation and $R_{0}$ of $\rho_{2}$ at $x=0.7$]{\includegraphics[height=5cms,width=5.5cms]{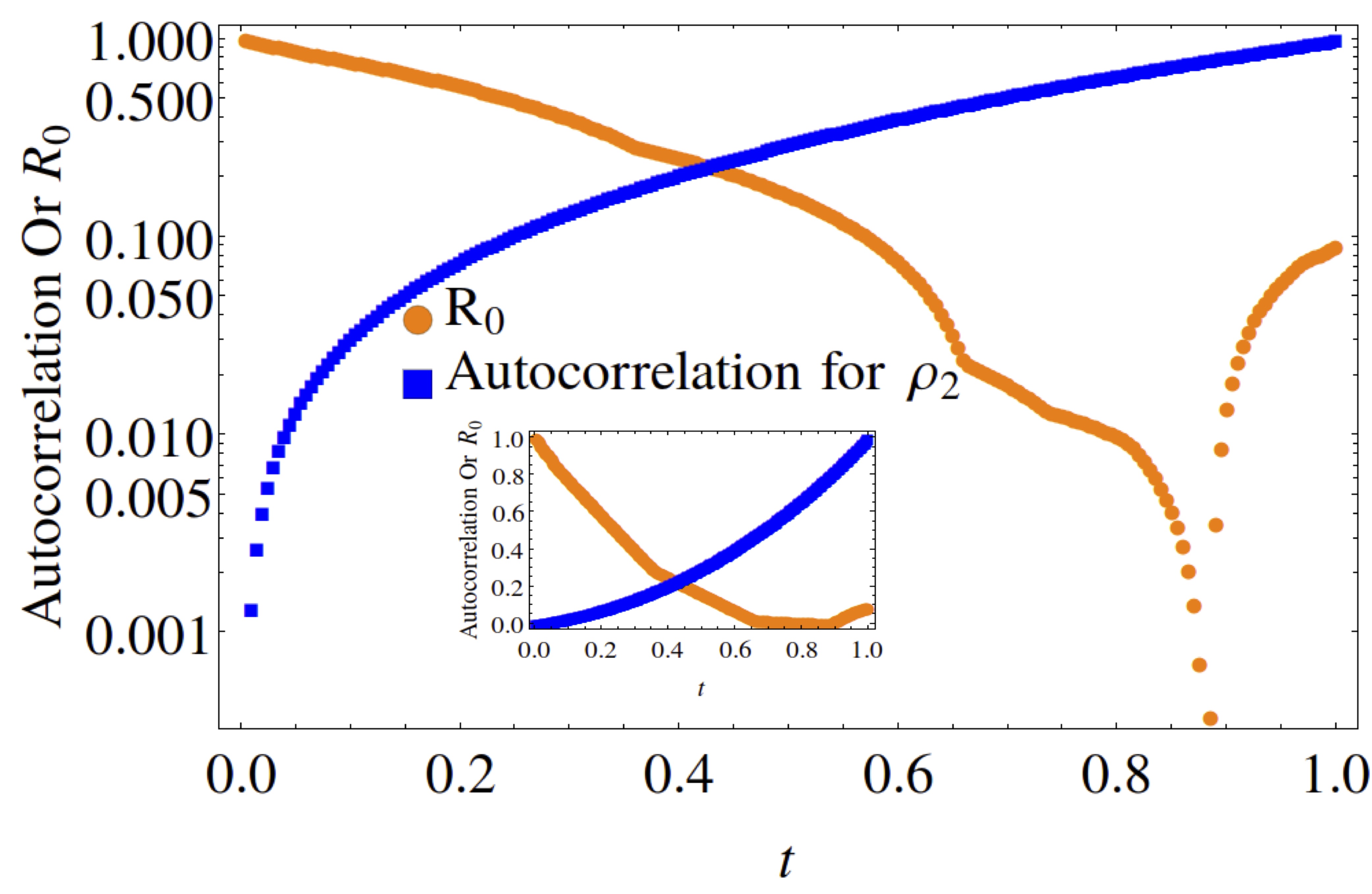}}
      \end{tabular}
        \caption{Comparison of normalized autocorrelation and $R_{0}$ at $x=0.7$. Outset is in log scale and inset is in original scale. \label{fig0p7}}
\end{figure}

{\color{black}{Note, the dynamics of $\rho_1$ as shown in Figure \ref{fig0p7} matches those for $x=0.3,0.5,0.6$. However, the species $\rho_2$ is mostly created by the conversion of the species $\rho_1$. These 5 figures clearly indicate that only for the symmetric case $x=0.5$, the time dynamical evolution of the reproductive number for an epidemic model matches the average energy dissipation rate of individual variables (expressed as autocorrelation functions), not otherwise. This is not unexpected as the point $x=0.5$ (spatial scale $0<x<1$) represents the point of dynamical equilibrium between two diffusing species, that also represents infection flux equilibrium between susceptible-infected-recovered species in an epidemic model. In other words, a fair quantitative comparison between the $R_0$ versus the D-D model is only ensured at $x=0.5$.}}

\section{Conclusions}

Clearly, a comparison of the dynamical variable $R_0(t)$, motivated by the 
epidemiological literature, with the autocorrelation function reveals the 
richness of the dynamics of a reaction-diffusion system which offers an option of interpolating the results from the epidemic model into the double-diffusion domain, in the process providing a closed form solution of the latter that has remained elusive thus far. Comparing the time evolution 
of $R_0$ with the autocorrelation function gives the information of the 
origin of the observed abundance of different species in a reaction-diffusion 
system as explained in Figures \ref{fig0p2} to \ref{fig0p7}.
The analogy is strictly restricted to the spatially symmetric ($x=0.5$) conformation though, a point of dynamical equilibrium between two (or multiple) diffusing species, an analogy with the stationary state fixed point of an epidemic model in dynamical equilibrium.

Therefore, the introduction of the epidemiologically motivated quantity $R_0(t)$ into the studies of the reaction-diffusion systems can play a crucial role in understanding such systems in more depth. Since this interpolation between two unrelated disciplines only uses the mathematical similarity between two (or multiple) reaction-diffusion species, expressed as double-diffusion in material science, as compared to infection rate growth in epidemiology, the approach is generic enough to be applied to all coupled reaction-diffusion models. At the point of symmetry ($x=0.5$ in our model), both quantities ($R_0$ and autocorrelation) will asymptotically match their values with evolving time allowing for a closed form mapped (from mathematical biology) solution of the R-D model. As a comparison with the numerical solution confirms close convergence with the approximate mapped solution (based on the $R_0(t)$ formula as a descriptor of the correlation strength of the diffusing variables), the solution provides handle to studies analyzing higher order perturbations and relevant bifurcations, also including stochastic terms. Future studies involving calculation of correlated superconducting fluxes would be presented using the same method.

\section*{Acknowledgment}
The authors gratefully acknowledge partial financial support from the H2020-MSCA-RISE-2016 program, grant no. 734485, entitled \enquote{Fracture Across Scales and Materials, Processes and Disciplines (FRAMED)}.

\end{document}